# Ion velocity distribution functions in argon and helium discharges: detailed comparison of numerical simulation results and experimental data

Huihui Wang[1,2], Vladimir S Sukhomlinov[3,4], Igor D Kaganovich[2], Alexander S Mustafaev[5].

[1]School of Physical Electronics, University of Electronic Science and Technology of China, Chengdu, 610054, China

[2]Princeton Plasma Physics Laboratory, Princeton, NJ 08543, USA

[3]Department of Physics, St. Petersburg State University, St. Petersburg, 198504, Russia

[4]ITMO University, St. Petersburg, 197101, Russia

[5]Department of General and Technical Physics, National Mineral-Resource University, St. Petersburg, 199106, Russia

E-mail: huihuiwang@uestc.edu.cn; ikaganov@pppl.gov

**Abstract**

Using Monte Carlo Collision (MCC) method, we have performed simulations of ion velocity distribution functions (IVDF) taking into account both elastic collisions and charge exchange collisions of ions with atoms in uniform electric fields for argon and helium background gases. The simulation results are verified by comparison with the experiment data of the ion mobilities and the ion transverse diffusion coefficients in argon and helium. The recently published experimental data for the first seven coefficients of the Legendre polynomial expansion of the ion energy and angular distribution functions are used to validate simulation results for IVDF. Good agreement between measured and simulated IVDFs shows that the developed simulation model can be used for accurate calculations of IVDFs.

**Keywords:** ion velocity distribution function, ion-atom angular differential cross section, Monte Carlo collision method

## 1. Introduction

Detailed knowledge of the ion velocity distribution functions (IVDF) is important for many applications [1-3], particularly for plasma-material surface interactions. Very simplified models that assume IVDF as a shifted Maxwellian velocity distribution function, are often used, e.g. for studies of dusty plasma [4-6], notwithstanding the fact that this simple model is not sufficient and simulated IVDFs are far from shifted Maxwellian velocity distribution functions.

Often, a treatment for IVDF considers only charge exchange collisions. For this approach, the simplest approximation uses a constant charge exchange collision frequency, so-called Bhatnagar-Gross-Krook (BGK) model [7]. It is easy to obtain an analytic expression for IVDF with the BGK model. However, the charge exchange cross section depends on the velocity weakly. This means that BGK model is not accurate for simulations of IVDF. In Ref.[8], Else D et al. carried out numerical solutions of IVDF for a constant charge exchange cross section to compare results with those obtained using the BGK model, and showed that the BGK model is not accurate in the limit of strong electric field.

Assuming a constant charge exchange cross section, an analytic solution for the IVDF was derived without taking into account the atom thermal temperature in Ref.[9] which is approximate in the limit of strong electric field, and a numerical solution for the IVDF taking into account the atom thermal temperature for a general value of the electric field was obtained by Lampe M et al. [10]. The recent study of IVDF is performed by Mustafaev A, et al., in which the analytical calculation of the IVDF was performed taking into account the atom thermal temperature [11].

Although the charge exchange collisions dominate the ion-atom collisions, the elastic collisions also affect IVDF, especially for the direction transverse to the electric field. Notwithstanding this fact, we are not aware of any publications studying IVDFs (including in DC - discharge), where both charge exchange collisions and scattering in polarizing potential are both taken into account. Therefore, in our previous publication [12], we have developed an approximate numerical model of angular differential cross sections for both elastic collisions and charge exchange collisions for simulations of the IVDFs in helium discharges, and have shown that associated errors in conventional approach where only charge exchange collisions are taken into account.

Recently, IVDFs are measured by Ref. [11, 15] making use of a planar one-sided probe [13-14]. The experimental measurements of IVDF allow for careful benchmarking of simulations and collision data (elastic and charge-exchange collision angular differential cross sections), which are necessary for accurate simulations of IVDFs. In this paper, our previously developed approximations for elastic and charge-exchange collision angular differential cross sections [12] were used for IVDF simulations in helium and argon, and the simulated IVDF are compared with the experimental data. The Numerical model is described in Sec.2, comparison of simulation results and experiment data is carried out in Sec.3, and conclusion is made in Sec.4.

## 2. Description of Monte Carlo Collision method for ion-atom collisions

In this section, we describe the Monte-Carlo Collisions (MCC) method applied for IVDF calculations. The detailed description can be found in our previous publication for helium [12]. Here, the method was also developed for argon. The ion-atom angular differential scattering cross section of both elastic collisions and charge exchange is approximated in the following form

$$\sigma_\theta(\varepsilon, \theta) = \frac{A(\varepsilon)}{[1-\cos\theta+a(\varepsilon)]^{1.25}} + \frac{A(\varepsilon)}{[1+\cos\theta+b(\varepsilon)]^{1.25}}, \quad (1)$$

where $\varepsilon$ is the relative translational energy in eV of ion in ion-atom center of mass reference frame ($\varepsilon$ is about 0.5 times of the ion energy) and $\theta$ is the scattering angle. Using this angular differential cross section, the total cross section, $\sigma_t$, the momentum transfer cross section, $\sigma_m$, and the viscosity cross section, $\sigma_v$, can be calculated analytically as given by expressions in Eq.(2), Eq.(3) and Eq.(4), respectively [12].

The functions $A(\varepsilon)$, $a(\varepsilon)$, $b(\varepsilon)$ are parameters of the model, which can be determined from known cross section $\sigma_t$, $\sigma_m$, and $\sigma_v$ (see Table 1). The numerical solution method of Eqs. (2-4) for functions $A(\varepsilon)$, $a(\varepsilon)$, $b(\varepsilon)$ is given in our previous publication [12]. The approximation formulas for $\sigma_{total}$, $\sigma_m$, and $\sigma_v$ are presented in Table 1.

In Table 1 data for argon gas are obtained using the empirical formula $\sigma_v(\varepsilon)$ [16] and the experimental data for $\sigma_m(\varepsilon)$; a quantum mechanical calculation for $\sigma_t(\varepsilon)$ are taken from Ref. [17]; data for helium gas are the proposed in Ref. [12] approximation of quantum mechanical calculations for $\sigma_t(\varepsilon)$ and the experimental data $\sigma_v(\varepsilon)$, $\sigma_m(\varepsilon)$ are taken also from Ref. [12].

Using Eq.(1), values of $A(\varepsilon)$, $a(\varepsilon)$, $b(\varepsilon)$, the angular differential cross sections are calculated and compared with the experimental data for angular differential cross sections of scattering of ions in its own gas for Ar$^+$+Ar [18] and He$^+$+He [19] systems, which are depicted in figure 1 showing an approximate agreement.

$$\sigma_t(\varepsilon) = 2\pi \int_0^\pi \sigma_\theta(\varepsilon,\theta) \sin\theta \, d\theta = 8\pi A \left[\frac{1}{a^{0.25}} - \frac{1}{(2+a)^{0.25}} + \frac{1}{b^{0.25}} - \frac{1}{(2+b)^{0.25}}\right]. \quad (2)$$

$$\sigma_m(\varepsilon) = 2\pi \int_0^\pi \sigma_\theta(\varepsilon,\theta)(1-\cos\theta) \sin\theta \, d\theta = 8\pi A \left[\frac{a}{(2+a)^{0.25}} - \frac{4a^{0.75}}{3} + \frac{(2+a)^{0.75}}{3} - \frac{4(2+b)^{0.75}}{3} + \frac{2}{b^{0.25}} + \frac{4b^{0.75}}{3}\right]. \quad (3)$$

$$\sigma_v(\varepsilon) = 2\pi \int_0^\pi \sigma_\theta(\varepsilon,\theta)(1-\cos^2\theta) \sin\theta \, d\theta = 8\pi A \left[\frac{2(2+a)^{0.75}}{3} - \frac{(2+a)^{1.75}}{7} - \frac{8a^{0.75}}{3} + \frac{5a(2+a)^{0.75}}{3} - \frac{32a^{1.75}}{21} + \frac{2(2+b)^{0.75}}{3} - \frac{(2+b)^{1.75}}{7} - \frac{8b^{0.75}}{3} + \frac{5b(2+b)^{0.75}}{3} - \frac{32b^{1.75}}{21}\right]. \quad (4)$$

Table 1. The approximation formulas for $\sigma_t$, $\sigma_m$, $\sigma_v$

|  | Argon | Helium |
| --- | --- | --- |
| $\sigma_m(\varepsilon), m^2$ | $1.15 \times 10^{-18} \times [1 + 0.015/(2\varepsilon)]^{0.6}(2\varepsilon)^{-0.1}$ [16] | $5.58 \times 10^{-19} \times [1 - 0.0557\ln(2\varepsilon)]^2 [1 + 0.0006\varepsilon^{-1.5}]$ [12] |
| $\sigma_v(\varepsilon), m^2$ | $\frac{2}{3} \times \left[\frac{2\times 10^{-19}}{(2\varepsilon)^{0.5}\times(1+2\varepsilon)} + \frac{3\times 10^{-19}\times 2\varepsilon}{(1+2\varepsilon/3)^2}\right]$ [16] | $\frac{\sigma_m(\varepsilon)}{1.5(1+\varepsilon^{1.1})}$ [12] |
| $\sigma_t(\varepsilon), m^2$ | $7.78 \times 10^{-18}/\varepsilon^{0.335}$ [17] | $\sigma_m(\varepsilon)[1 + \varepsilon^{-0.2}]$ [12] |

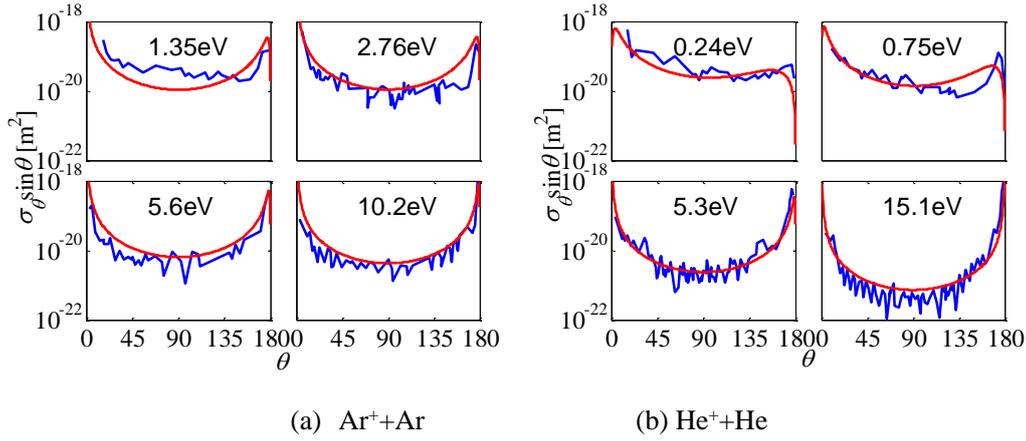

(a) Ar$^+$+Ar  (b) He$^+$+He

Figure 1. The angular differential cross section in Ar$^+$+Ar, and He$^+$+He systems. The blue curves show the experimental data [18-19] and the red curve is approximation given by Eq.(1).

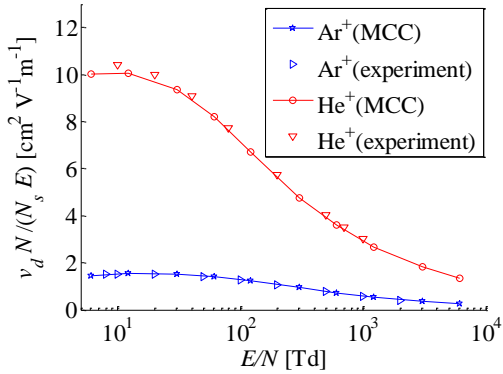

Figure 2. The reduced mobility as a function of reduced electric field

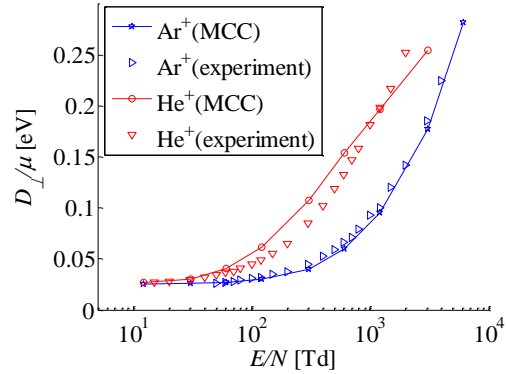

Figure 3. The transverse diffusion as a function of reduced electric field

It should be noted that Phelps proposed model cross section for the description of experimental data for Ar$^+$+Ar, He$^+$+He and H$^+$+H in Ref. [20]. However the proposed fits for cross sections assume symmetry regarding transformation $\theta \rightarrow \pi - \theta$. This approximation does not fully describe the experimental data at small energies, see Fig. 1b.

In simulations, the actual scattering ion-atom collision is divided to two parts: one part describes the small-angle scattering $\sigma_1(\varepsilon, \theta) = A/(1 - \cos\theta + a)^{1.25}$, and the other - the scattering on the angle of about $\pi$, $\sigma_2(\varepsilon, \theta) = A/(1 + \cos\theta + b)^{1.25}$. The scattering angles ($\theta_1$ and $\theta_2$) in MCC simulations are therefore controlled by uniformly distributed random numbers between 0 and 1 ($R_1$ and $R_2$) according to Eq.(5) and Eq.(6) for these two parts, respectively [12].

$$\cos\theta_1 = 1 + a - \{a^{-0.25} - R_1[a^{-0.25} - (2+a)^{-0.25}]\}^{-4}. \quad (5)$$

$$\cos\theta_2 = -(1+b) + \{(2+b)^{-0.25} + R_2[b^{-0.25} - (2+b)^{-0.25}]\}^{-4}. \quad (6)$$

MCC particle simulations were performed with the values of $a$, $b$, $A$, and making use of Eqs.(5-6) [12]. If $\sigma_v$ is negligible, then $A \rightarrow 0$, $a \rightarrow 0$, and $b \rightarrow 0$, which makes $\cos\theta_1 = 1$ for elastic collisions, and $\cos\theta_2 = -1$ for charge exchange collisions. Therefore this collision process is reduced to the only charge exchange collisions of the scattering on $\pi$-angle in the center mass reference frame [10].

The MCC method in this paper ($\sigma_m$, $\sigma_v$ and $\sigma_t$ are taken from table 1) is verified by comparing simulated results for the mobility and the transverse diffusion with the experimental data reported in Refs. [21-24] as shown in figure 2 and figure 3, at the discharge condition of 294K gas temperature and 0.1Torr gas pressure, where $v_d$ is the ion drift velocity, $E$ is the electric field, $N$ is the gas density, $N_s$=2.6868×10$^{19}$ cm$^{-3}$ is the standard gas number density, $\mu$ is the mobility ($v_d/E$), and $D_\perp$ is the transverse diffusion coefficient. Note that the ion mobility is fully determined by the momentum transfer cross section, $\sigma_m$, whereas the transverse diffusion coefficient is mostly function of the viscosity cross section, $\sigma_v$. Besides this validation, the MCC code has been also benchmarked with another well-used PIC code EDIPIC [25].

## 3 Ion velocity distribution functions and comparison to experiment

In this section, following conditons are used $T$=450K, $p$=0.2torr and $E/p$=9V/(cm•torr) for the argon discharge, and $T$=600K, $p$=0.2torr and $E/p$=20V/(cm•torr) for the helium discharge. IVDFs were measured by a flat one-sided probe making use of the second derivative of the current relative to biased voltage for different orientations of the probe and by applying decomposition of angular dependence in the Legendre polynomials [11, 15]. Number of polynomial coefficients equals to the number of probe angular orientations. Therefore, IVDF is represented as a finite summ of Legendre polynomials $\sum_{k=0}^{N} F_n(\varepsilon_{ion}) P_k(cos\theta)$, where $\varepsilon_{ion}$ is the ion energy, $F_n(\varepsilon_{ion})$ is the coefficient for a Legendre polynomials, $\theta$ is the angle between the ion velocity direction and the electric field direction, and $P_k(cos\theta)$ is the Legendre polynomial of order $k$. The more anisotropic IVDF is, the more coefficients have to be used for correct representation. For nearly isotropic IVDF only zeroth term can be used; for very anisotropic IVDF pointing into only one direction (IVDF is delta- function of angle) infinite number of terms have to be used. Criterion for sufficient number of polynomials is that the IVDF calculated in ($N$+1)-approximation is very close to the IVDF calculated in N-approximation. Typically, the high energy tail of IVDF is more anisotropic than the bulk of IVDF, see Fig.4.

In order to compare simulated IVDF to the experiment data, we perform the Legendre expansion of the IVDF expressed as energy and angle distribution function $F(\varepsilon_{ion}, \theta)$ normalized according to Eq.(7), as performed for experimental data

$$\int_0^{+\infty} \int_0^{\pi} F(\varepsilon_{ion}, \theta) \sin\theta \, d\theta \, d\varepsilon = 1. \quad (7)$$

For the Legendre expansion

$$F(\varepsilon_{ion}, \theta) = \sum_{n=0}^{+\infty} F_n(\varepsilon_{ion}) P_n(cos\theta), \quad (8)$$

its coefficients are given by

$$F_n(\varepsilon_{ion}) = \frac{2n+1}{2} \int_0^{\pi} F(\varepsilon_{ion}, \theta) P_n \sin\theta \, d\theta, \quad (9)$$

and the ion energy distribution function *IEDF* is

$$IEDF(\varepsilon_{ion}) = \int_0^{\pi} F(\varepsilon_{ion}, \theta) \sin\theta \, d\theta \equiv 2F_0 \quad (10)$$

The 1st, 2nd, 4th, 6th order Legendre expansions are presented in figure 4. For argon, the 2nd order expansion is close to 6th order expansion for $\varepsilon$ =0.03eV, while there is a noticable difference between the 2nd order expansion and the 6th order expansion for $\varepsilon$ =0.09 eV. The similar phenomenon is also found for helium. Apparently, for the electron energy 0.5 eV the 4th order expansion is significantly different from the 6th order expansion for He (see Fig. 4b, indicated by the arrows). This means that the angluar IVDFs at low ion energies are more isotropic than the angluar IVDFs at high ion energies.

Combining Eq. (7) and the normalization condition Eq.(11)

$$\int_0^{+\infty} \int_0^{\pi} f(v, \theta) 2\pi v^2 \sin\theta \, d\theta \, dv = 1. \quad (11)$$

IVDF could be obtained from $F(\varepsilon, \theta)$.

$$f(v,\theta) = \frac{F(\varepsilon_{ion}, \theta)}{2\pi v} \frac{M}{e} = \frac{F\left(\frac{M(v_x^2+v_y^2+v_z^2)}{2e}, \arctan\frac{\sqrt{v_y^2+v_z^2}}{v_x}\right)}{2\pi v} \frac{M}{e}, \quad (12)$$

where $M$ is the ion mass, $v$ is the ion speed, and $e$ is the elementary charge. Assuming $x$ direction is along the electric filed, we only focus on the two dimensional IVDF $f(v_x, v_y)$ because IVDF is axisymmetric in $y$ and $z$ directions.

$$f(v_x, v_y) = \int_{-\infty}^{\infty} f(v_x, v_y, v_z) dv_z \quad (13)$$

Equations (12-13) relates $F(\varepsilon_{ion}, \theta)$ with $f(v_x, v_y)$.

Making use of Eqs. (8), (12) and (13), two dimensional IVDF of the 6-order Legendre expansion is calculated and is shown in figure 5, where $v_T$ is the thermal velocity of the according to the gas temperature. Figure 5 presents IVDF of the 6 order Legendre expansion are consistent with IVDF of full calculation, which shows the accuracy of the 6 order Legendre expansion is high enough for these discharge conditions.

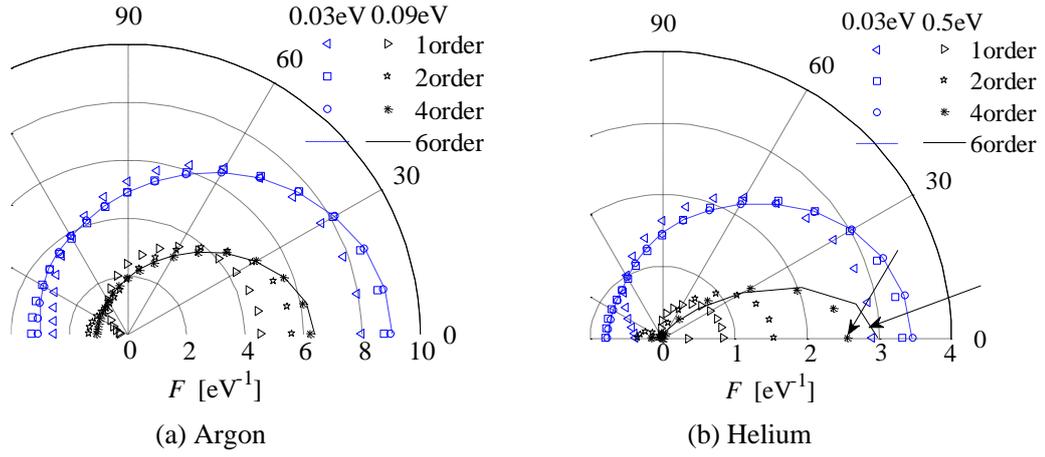

(a) Argon  (b) Helium

Figure 4. Angular distribution of different order Legendre expansions.

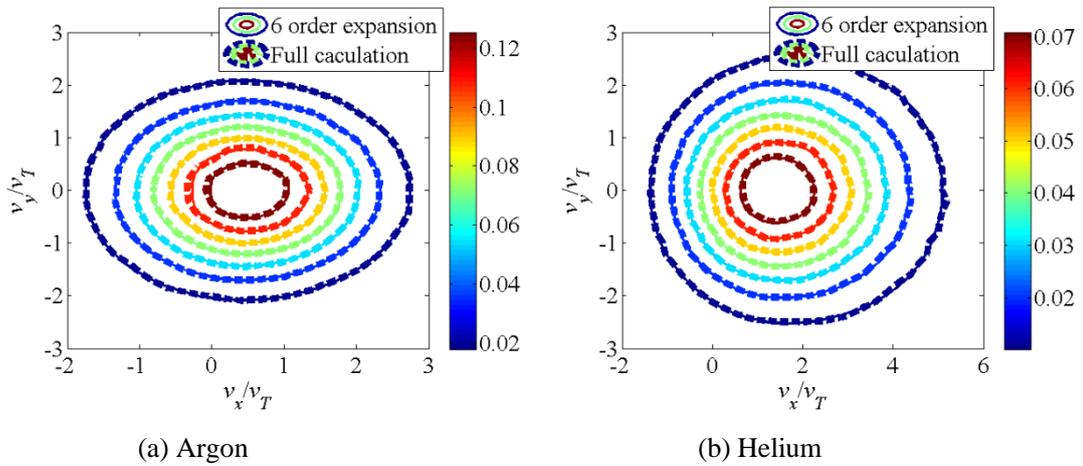

(a) Argon  (b) Helium

Figure 5. The accuracy of 6 order Legendre expansion for $f(v_x/v_T, v_y/v_T)$.

In the following, IVDF obtained in MCC simulations is compared to the experiment data. In the experiment, the Legendre expansion coefficients of $F(\varepsilon_{ion}, \theta)$ are affected by the instrument function $A$. The measured distribution function $F_{measure}$ is the convolusion of the real distribution function $F_{real}$ and the instrument function $A$ [11, 26]. This convolution operation is given by Eqs. (14-15), and its application leads to a decrease $F(\varepsilon_{ion})$ near maximum and increase near $\varepsilon_{ion} = 0$:

$$F_{measure}(\varepsilon_{ion}) = \frac{\sqrt{2}}{2.221\delta}\int_{-\infty}^{\infty} F_{real}(\varepsilon_{ion} - \varepsilon')A\left(\frac{\sqrt{2}\varepsilon'}{\delta}\right)d\varepsilon' \quad (14)$$

$$A(z) = \begin{cases} \frac{8}{\pi}\int_{\frac{|z|}{2\sqrt{2}}}^{1}\sqrt{\frac{\left(u^2 - \frac{z^2}{8}\right)}{u}}(1-u)du, |z| \leq 2\sqrt{2} \\ 0, |z| > 2\sqrt{2} \end{cases} \quad (15)$$

where $\delta$ is the energy resolution step. Apparently, $F_{measure}$ approaches to $F_{real}$ if $\delta$ equals to 0. However, usually $\delta$ in the experimental conditions is not sufficiently small, because of limitations of the measurement technology. This means that applying the convolution operation is necessary for comparison with the experimental data. We take the MCC results as $F_{real}$, and use $\delta = 0.05$eV according to suggestion of Ref. [15].

The effect of the convolution operation is to average the distribution function over $[\varepsilon_{ion} - 2\delta, \varepsilon_{ion} + 2\delta]$ with the weight function given by Eq.(15). The Legendre polynomial expansion coefficients before and after the convolution for argon and helium are shown in figure 6. As expected, the peaks of $F_n(\varepsilon_{ion})$ decrease and the values of $F_n(0)$ at the zero ion energy increase after convolution, as shown in figure 6. And the effect of convolution in figure 6(a) is more significant than that in figure 6(b) because the argon discharge has a sharper energy distribution.

$F_n(\varepsilon_{ion})$ terms simulated by MCC after applying convolution are compared with the experimental data for argon and helium and are shown in figure 6(a) and 6(b), respectively. Figure 6(a) shows a good agreement between the MCC and experimental data for n=0. Although there are some errors for the higher order Legendre coefficients, the effect of these errors on the total IVDF is small as evident in figure 7(a), because the Legendre coefficients decreases quickly with the Legendre order n under this discharge condition. Experimental error for n=0 is estimated at 10% and increases with n because of signal reduction. Figure 6(b) shows the same comparison for the helium discharge. Even for n=0, there are some errors between MCC results and experimental data. The errors become more pronounced for higher ion energies. These errors are examined in contour plots of the total IVDF depicted in figure 7.

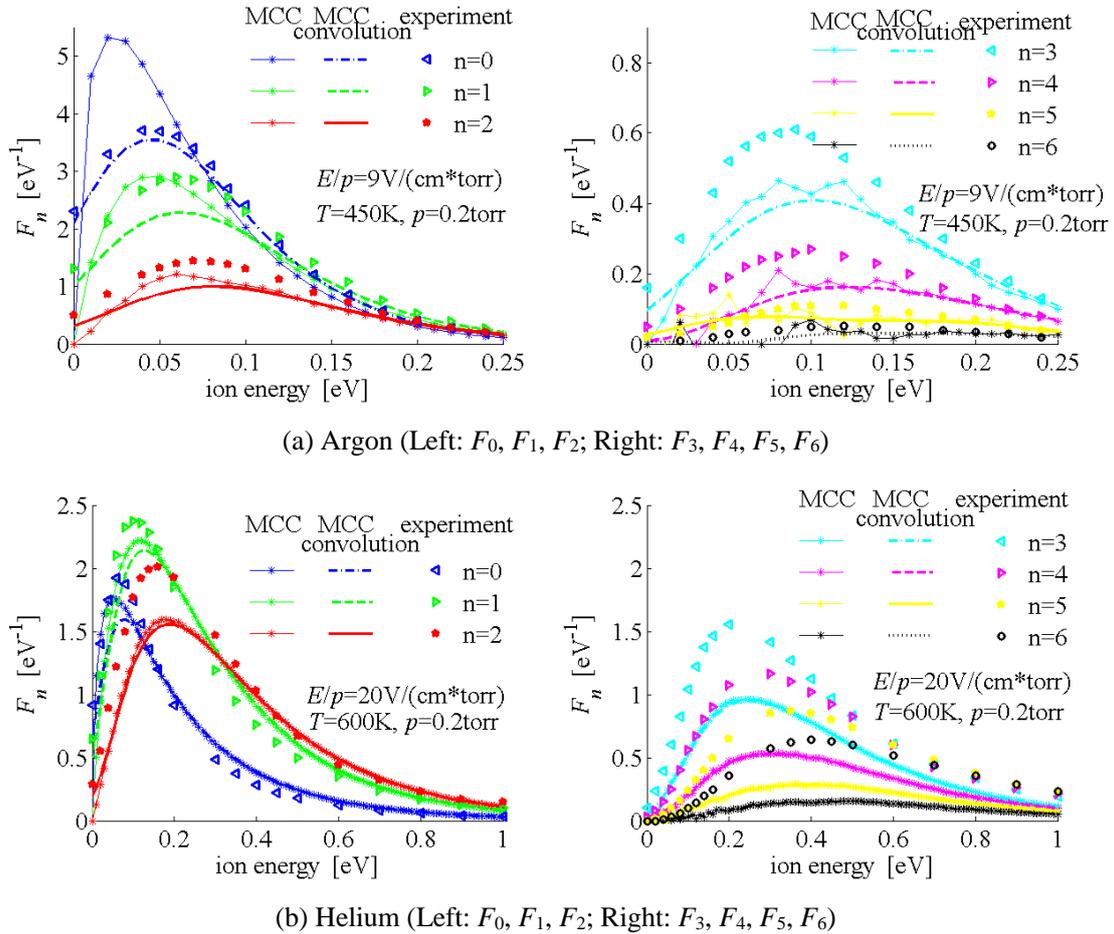

(a) Argon (Left: $F_0$, $F_1$, $F_2$; Right: $F_3$, $F_4$, $F_5$, $F_6$)

(b) Helium (Left: $F_0$, $F_1$, $F_2$; Right: $F_3$, $F_4$, $F_5$, $F_6$)

Figure 6. Comparison of calculations with experimental data for the energy dependence of the Legendre polynomials expansion coefficients for distribution function (a) Argon, (b) Helium.

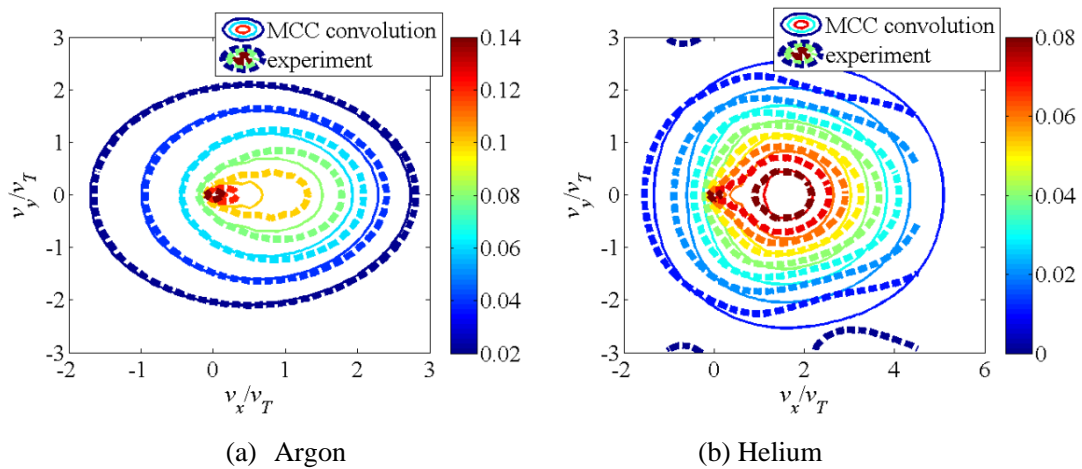

(a) Argon  (b) Helium

Figure 7. $f(v_x/v_T, v_y/v_T)$ obtained with the 6$^{th}$ order expansion after the convolution operation.

Combining Eqs.(8), (12) and (13), IVDFs are calculated from the result of the 6-order Legendre polynomial expansion after convolution, which is presented in Figure 7. Figure 7(a) shows a good agreement between MCC and experiments for argon, while there are some errors for high ion velocity in Figure 7(b) for helium. Namely, the experimentally determined IVDF is more anisotropic at high energies.

Furthermore, it is worth noting that the 6-order Legendre expansion may not be sufficient if degree of IVDF anisotropy is large, e.g. for a higher electric field or a lower pressure, and a higher order Legendre expansion is necessary in this case.

# 4 Conclusion

In summary, we have simulated ion velocity distribution functions of $Ar^+, He^+$ in plasmas of glow discharge in argon and helium, respectively. For simulations, we have used approximations for charge exchange and scattering angular differential cross sections developed earlier in Ref. [12]. The proposed model describes well experimental data for angular differential cross sections for $Ar^+ + Ar, He^+ + He$ [18-19]. Parametrization of angular differential cross sections uses available data for the momentum transfer, viscosity and total cross sections; latter cross sections are well verified using available experimental data of mobility and diffusion.

Comparison of simulated IVDFs with the data measured by a flat probe showed good agreement for $Ar^+ + Ar$ and reasonable agreement for $He^+ + He$. The difference between measured and simulated IVDFs may be attributed to insufficient resolution of measured IVDF, because only seven polynomials were used for strongly anisotropic IVDF.

Good agreements between measured and simulated IVDFs show that the developed siumulation model can be used for accurate calculations of IVDFs.

Acknowledgment: We are thankful to Alex Khrabrov for fruitful discussions and benchmarking of the Monte Carlo code used in the paper with EDIPIC code and Predrag Kristic for valuable discussions on cross sections. The work of H. Wang was supported by China Scholarship Council and the Fundamental Research Funds for the Central Universities ZYGX2014J040, Igor D. Kaganovich was supported by U.S. Department of Energy,

# Reference

[1] http://science.energy.gov/~/media/fes/pdf/about/Low_temp_plasma_report_march_2008.pdf

[2] Morfill G EandIvlev A V 2009 Complex plasmas: An interdisciplinary research field *Reviews of Modern Physics* **81** 1353

[3] Fortov V E, Ivlev A V, Khrapak S A, Khrapak A G and Morfill G E 2005 Complex (dusty) plasmas: Current status, open issues, perspectives *Physics Reports* **421** 1

[4] Lampe M, Joyce G and Ganguli G 2000 Interactions between dust grains in a dusty plasma *Physics of Plasmas* **7** 3851

[5] Lapenta G 2000 Linear theory of plasma wakes *Physical Review E* **62** 1175

[6] Khrapak S A, Ivlev A V, Zhdanov S K and Morfill G E 2005 Hybrid approach to the ion drag force *Physics of Plasmas* **12** 042308

[7] Bhatnagar P L, Gross E R and Krook M 1954 AModel for Collision Processes in Gases. I. Small Amplitude Processes in Charged and Neutral One-Component Systems *Physical Review* **94** 511

[8] Else D, Kompaneets R and Vladimirov S V 2009 On the reliability of the Bhatnagar-Gross-Krook collision model in weakly ionized plasmas *Physics of Plasmas* **16** 062106

[9] Smirnov B M 2007 Plasma Processes and Plasma Kinetics (Wiley-VCH) 261

[10] Lampe M, Röcker T B, Joyce G, Zhdanov S K, Ivlev A V and Morfill G E 2012Ion distribution function in a plasma with uniform electric field *Physics of Plasmas* **19** 113703

[11] Mustafaev A S, Sukhomlinov V S and Ainov M A 2015 Experimental and theoretical determination of the strongly anisotropic velocity distribution functions of ions in the intrinsic gas plasma in strong field *Technical Physics* **60** 1778

[12] Wang H, Sukhomlinov V, Kaganovich I D and Mustafaev A S 2016 Simulations of Ion Velocity Distribution Functions Taking into Account Both Elastic and Charge Exchange Collisions *Plasma Sources Sci. Technol.* (accepted)

[13] Mustafaev A S and Grabovskii A Yu 2012 Probe Diagnostics of Anisotropic Electrons Distribution

Function in Plasma *High Temperature* **50** 785-805

[14] Mustafaev A S 2001 Dynamics of electron beams in plasmas *Technical Physics* **46** 111

[15] Sukhomlinov V, Mustafaev A S, Grabovskii A and Ainov M 2015 Ion velocity distribution function in the plasma of its own gas *42nd EPS Conference on Plasma Physics* P5.168

[16] Phelps A V 1994 The application of scattering cross sections to ion flux models in discharge sheaths *Journal of Applied Physics* **76** 747

[17] Barata J A S 2007 Integral and differential elastic collision cross-sections for low-energy $Ar^+$ ions with neutral Ar atoms *Nuclear Instruments and Methods in Physics Research A* **580** 14-17

[18] Vestal M L, Blakley C R and Futrell J H 1978 Crossed-beam measurements of differential cross sections for elastic scattering and charge exchange in low-energy $Ar^+$-Ar collisions *Physical Review A* **7** 1337-1342

[19] Vestal M L, Blakley C R and Futrell J H 1978 Crossed-beam measurements of differential cross. sections for elastic scattering and charge exchange in low-energy $He^+$-He collisions *Physical Review A* **17** 1321

[20] Phelps A V 2007
http://fr.lxcat.net/notes/index.php?download=phelps3

[21] Ellis H W, Pai R Y and Mcdaniel E W 1976 Transport properties of gaseous ions over a wide energy range *Atomic Data And Nuclear Data Tables* **17** 177-210

[22] Helm H 1977 The cross section for symmetric charge exchange of $He^+$ in He at energies between 0.3 and 8eV J. *Phys. B: Atom. Molec. Phys.* **10** 3683

[23] Stefansson T, Berge T, Lausund R and Skullerud H R 1988 Measurements of transport coefficients for lithium ions in argon and helium ions in helium with a drift-tube mass spectrometer *J. Phys. D* **21** 1359

[24] Stefansson T and Skullerud H R 1999 Measurements of the ratio between the transverse diffusion coefficient and the mobility for argon ions in argon *J. Phys. B: At. Mol. Opt. Phys.* **32** 1057-1066

[25] Sydorenko D 2006 Particle-in-cell simulations of electron dynamics in low pressure discharges with magnetic fields *PhD Thesis* University of Saskatchewan

[26] Volkova L M, Demidov V I, Kolokolov N B, Kralkina E M 1984 Comparison, based on instrumental functions, of different probe methods for measuring the energy-distribution of electrons in a plasma *High Temperature* **22** 612